\documentclass[pre,amsmath, twocolumn, showpacs, superscriptaddress]{revtex4}
\usepackage{graphicx}

\begin{document}

\title{An illustrative example of the relationship between dissipation and relative entropy}

\author{Jordan Horowitz}

\affiliation{Department of Physics, University of Maryland, College Park, MD 20742 USA}

\author{Christopher Jarzynski}

\affiliation{Department of Chemistry and Biochemistry, and Institute for Physical Science and Technology,\\
University of Maryland, College Park, MD 20742 USA}

\date{\today}

\begin{abstract}
Kawai, Parrondo, and Van den Broeck [{\it Phys.\ Rev.\ Lett.}\ {\bf 98}, 080602 (2007)] have recently established a quantitative relationship between dissipated work and a microscopic, information-theoretic measure of irreversibility.
We illustrate this result using the exactly solvable system of a Brownian particle in a dragged harmonic trap.
\end{abstract}


\maketitle

\section{Introduction}

During a thermodynamic process in which a system, in contact with a thermal reservoir, evolves from one state of thermal equilibrium $A$, to another $B$, the average work performed on the system must exceed the free energy difference: $\langle W\rangle \ge\Delta F = F_{B}-F_{A}$~\cite{Landau5}.
The work in excess of $\Delta F$, i.e.\  the {\it dissipated work} $\left\langle W_{\textrm{diss}}\right\rangle = \langle W\rangle - \Delta F$, quantifies the amount of energy irretrievably lost to the surrounding thermal environment.
Recently, Kawai, Parrondo, and Van den Broeck~\cite{Kawai2007} have related $\left\langle W_{\textrm{diss}}\right\rangle$ to another measure of irreversibility, roughly speaking the distinction between the forward and reverse directions of the ``arrow of time''.
Specifically, they have shown that
\begin{equation}
\label{eq:kpv2}
\beta\langle W_{\textrm{diss}}\rangle \ge D(\rho_{F}||\rho_{R}),
 \end{equation}
where $\rho_F$ and $\rho_R$ are time-dependent phase-space densities describing the evolution of the system from $A$ to $B$, and from $B$ to $A$, respectively; and
$D(\rho||\rho^\prime) = \int \rho \ln(\rho/\rho^\prime)$ denotes the {\it relative entropy} \cite{Cover}, a measure of the distinguishability between two distributions. 
When the system in question evolves deterministically, under Hamilton's equations, then Eq.~\ref{eq:kpv2} is an equality.  
However, if the system description is coarse-grained, or explicitly stochastic, then the relative entropy only provides a lower bound on the dissipated work.

Connections between dissipation and temporal asymmetry similar to Eq.\ \ref{eq:kpv2} have also been established by other authors.
Maes~\cite{Maes1999} and later Maes and Neto\v{c}n\' y~\cite{Maes2003} have obtained a correspondence between thermodynamic measures of dissipation and relative entropies between forward and reverse distributions in path space;
and Gaspard~\cite{Gaspard2004b} has connected this result to the dynamical randomness that characterizes nonequilibrium steady states.
In the context described in the previous paragraph, one of us~\cite{Jarzynski2006a} has derived a relation analogous to Eq.~\ref{eq:kpv2}, but expressed in terms of distributions in path space (see Eq.~\ref{eq:pathSpace}).

The goal of the present paper is to illustrate Eq.~\ref{eq:kpv2} using the simple model of a particle inside a one-dimensional, moving harmonic well.
This model is both analytically tractable and experimentally relevant~\cite{Mazonka1999,vanZon2004a, Garnier2005, Wang2005a, Wang2002, Carberry2004a, Douarche2006}.
After briefly reviewing the central result of Ref.~\cite{Kawai2007} in Section~\ref{sec:rev}, we introduce our model at the beginning of Section~\ref{sec:example}.
We then solve the model and illustrate Eq.~\ref{eq:kpv2} when the system evolves under Hamilton's equations (Section~\ref{sec:det}), and both overdamped and inertial Langevin dynamics (Sections~\ref{sec:od} and \ref{sec:full_lang}, respectively).
In Section~\ref{sec:two_time} we extend our analysis, and show that the bound in Eq.~\ref{eq:kpv2} can be improved by specifying the phase space density at two times, rather than one.
Gomez-Marin {\it et al.}\ \cite{Gomez-Marin2007,Gomez-Marin2008b} have also studied this model in the context of Eq.~\ref{eq:kpv2} and in Section \ref{sec:two_time} we briefly discuss the relationship between our results and theirs.

\section{Background}\label{sec:rev}

Consider a classical system with $N$ degrees of freedom, described by the coordinates ${\mathbf x} = \{x_{1},\dots,x_{N}\}$ and conjugate momenta ${\mathbf p} = \{p_{1},\dots,p_{N}\}$.  Let $z=({\mathbf x},{\mathbf p})$ denote a point in phase space, and let $H(z,\lambda)$ denote a parameter-dependent Hamiltonian for this system.
We assume that this Hamiltonian is time-reversal invariant: $H(z^{*},\lambda)=H(z,\lambda)$, where the asterisk denotes the reversal of momenta, ${\mathbf p}\to -{\mathbf p}$.

The term {\it thermodynamic process} will indicate a sequence of events whereby the system evolves in phase space as the external parameter $\lambda$, is varied according to an arbitrary schedule, or {\it protocol} $\lambda_t$ (also labeled as $\lambda(t)$), from an initial time $t=0$ to a final time $t=\tau$.
During this interval of time, the microscopic evolution of the system is specified by a trajectory $z_t$, or $z(t)$, and $\rho(z,t)$ will denote the time-dependent phase space density describing an ensemble of such trajectories.
As in Ref.~\cite{Kawai2007} we will explicitly consider two such processes: one defined by a {\it forward} protocol $\lambda_t^F$, during which the parameter is varied from an initial value $\lambda_0^F=A$ to a final value $\lambda_\tau^F=B$, the other defined by a {\it reverse} protocol $\lambda_t^R = \lambda_{\tau-t}^F$, from $\lambda_0^R=B$ to $\lambda_\tau^R=A$.

Prior to the start of either the forward or the reverse process ($t<0$), the system is brought to equilibrium by weak contact with a thermal reservoir at inverse temperature $\beta$.
As a result, the initial phase space density is a canonical distribution at $\lambda=A$ or $B$.
The system might subsequently be thermally isolated from the reservoir, or else it might remain in contact with the reservoir.
In the former case, Hamilton's equations govern its evolution from $t=0$ to $t=\tau$, while in the latter case we will use stochastic dynamics to model the random effects of the environment.
In either situation the {\it work} performed on the system during this interval of time is given by the following functional of the trajectory:
\begin{equation}\label{eq:work}
W\left[z_t\right] = \int_{0}^{\tau}{\mathrm d}t \, \dot{\lambda} \, \frac{\partial H}{\partial\lambda}(z_t,\lambda_t).
\end{equation}

For the case of Hamiltonian dynamics, the average dissipated work can be expressed as (see Ref.\ \cite{Kawai2007} for the details of the derivation)
\begin{eqnarray}\label{eq:kpv3}
\nonumber
\beta\langle W_{\mathrm{diss}} \rangle 
&=& \int{\mathrm d}z \, \rho_{F}(z,t)\, \ln\left[\frac{\rho_{F}(z,t)}{\rho_{R}(z^{*},\tau-t)}\right] \\
&=& D\left(\rho_{F}(z,t)||\rho_{R}(z^{*},\tau-t)\right).
\end{eqnarray}
Since relative entropy is a measure of the distinguishability of two probability distributions \cite{Cover}, Eq.~\ref{eq:kpv3} relates the dissipation of energy to the ease with which a process can be distinguished from its time-reversal~\cite{Kawai2007,Gomez-Marin2007,Gomez-Marin2008b, Gomez-Marin2008a}.

Eq.~\ref{eq:kpv3} is an {\it equality} because the underlying Hamiltonian dynamics are deterministic.
If the system remains in contact with the reservoir during the process, we instead obtain an inequality, Eq.~\ref{eq:kpv2}.
There are two ways to argue this, both of which make use of a key property of the relative entropy between two distributions, namely that it decreases when the distributions are projected onto a smaller set of variables~\cite{Cover}.
When the system remains in contact with a reservoir, then we can view the system and the reservoir as two sub-systems that together form a large, isolated ``super-system'' to which Eq.~\ref{eq:kpv3} applies:
the dissipated work gives the relative entropy between the two distributions in the full phase space.
Upon projecting out the reservoir variables, the relative entropy decreases, and we obtain Eq.~\ref{eq:kpv2}, as discussed in Ref.\ \cite{Kawai2007}.
Alternatively, we note that an equality analogous to Eq.~\ref{eq:kpv3} can be formulated for distributions in {\it path space}, rather than phase space~\cite{Jarzynski2006a}:
\begin{equation}
\label{eq:pathSpace}
\beta\langle W_{\textrm{diss}}\rangle= D({\cal P}_{F}(\gamma^F)||{\cal P}_{R}(\gamma^R)) .
\end{equation}
Here the trajectory $\gamma^F$ describes the evolution of the system during a given realization of the forward process, and ${\cal P}_{F}(\gamma^F)$ is the probability distribution of such trajectories;
$\gamma^R$ and ${\cal P}_R$ are defined in a similar manner for the reverse process.
Equation \ref{eq:pathSpace} holds for both deterministic and stochastic evolution.
If we now project from path space onto phase space, e.g.\ ${\cal P}_F(\gamma^F) \rightarrow \rho_F(z,t)$, then the relative entropy decreases, and we again obtain the inequality, Eq~\ref{eq:kpv2}. (References \cite{Gomez-Marin2007, Gomez-Marin2008b} refer to this projection procedure as ``coarse-graining in time''.)
In either case, the decrease of relative entropy has a simple interpretation:
when we discard microscopic information  -- either about the state of the reservoir, or about states of the system itself at times other than the specified instant, $t$ -- then we diminish our ability to distinguish between the forward and the reverse process.
This suggests that the more microscopic information we retain, the closer the value of $D$ is to its upper bound $\beta\langle W_{\rm diss}\rangle$.
We investigate this in Section~\ref{sec:two_time}, where we consider a generalization of Eq.~\ref{eq:kpv2} in which the microstate of the system is specified at two instants in time, rather than one.
References \cite{Kawai2007,Gomez-Marin2007, Gomez-Marin2008b, Gomez-Marin2008a} contain a related analysis specifically addressing the loss of microscopic information during coarse-graining and how this loss affects the value of the relative entropy.

\section{Particle In a Moving Harmonic Well}\label{sec:example}

In this section we illustrate Eq.~\ref{eq:kpv2} through an explicit calculation of the average dissipated work and relative entropy for a particle in a one-dimensional, moving harmonic well. We begin by specifying the model and obtaining useful preliminary results.  Then in section \ref{sec:det} we evaluate the average dissipated work and relative entropy for a system following Hamiltonian dynamics.  In sections \ref{sec:od} and \ref{sec:full_lang} we repeat the calculation with the system modeled by Langevin dynamics, in different limiting regimes.

Our system is a particle of mass $m$ trapped in a harmonic well with spring constant $k$:
\begin{equation}\label{eq:pot}
H(x,p,\lambda) = \frac{p^{2}}{2m} + \frac{k}{2}(x-\lambda)^{2} .
\end{equation}
We will consider processes during which the center of the well is moved either rightward or leftward at constant speed $u$.
These correspond to forward and reverse protocols,
$\lambda_F(t) = ut$ and $\lambda_R(t) = u(\tau-t)$.
Explicit expressions for the initial equilibrium densities are given by the following Gaussians:
\begin{subequations}
\label{eq:init_den}
\begin{align}
\rho_{F}(z,0) &= \frac{\beta}{2\pi}\sqrt{\frac{k}{m}}\exp\left[-\beta\left(\frac{p^{2}}{2m}+\frac{kx^{2}}{2}\right)\right], \\ 
\rho_{R}(z,0) &= \frac{\beta}{2\pi}\sqrt{\frac{k}{m}}\exp\left[-\beta\left(\frac{p^{2}}{2m}+\frac{k}{2}(x-u\tau)^{2}\right)\right].
\end{align}
\end{subequations}
For this system $\Delta F=0$ by translational symmetry, therefore for the forward process
\begin{equation}\label{eq:avg_work}
\langle W_{\textrm{diss}}\rangle = -uk\int_{0}^{\tau}{\mathrm d}t [\bar{x}_{F}(t)-ut],
\end{equation}
where $\bar{x}_{F}(t)$ is the average position during the forward process.

Since the time-dependent densities $\rho_F(z,t)$ and $\rho_R(z,t)$ will prove to be Gaussians for all the cases considered in this paper (see Sections~\ref{sec:det} - \ref{sec:full_lang}), it is useful here to establish uniform notation for the description of two-dimensional Gaussian distributions $f_G(z)=f_G(x,p)$.
Such a distribution is uniquely determined by the moments (means and covariances),
\begin{subequations}\label{eq:gaussian}
\begin{eqnarray}
\label{eq:x_mean}
\bar{x} & = &\int {\mathrm d}zf_{G}(z)x, \\ 
\label{eq:p_mean}
\bar{p} & = & \int {\mathrm d}zf_{G}(z)p, \\
\sigma_{x}^{2} & = & \int{\mathrm d}zf_{G}(z)(x-\bar{x})^2, \\ 
\sigma_{p}^{2} & = & \int{\mathrm d}zf_{G}(z)(p-\bar{p})^2, \\
\sigma_{xp} & = & \int{\mathrm d}zf_{G}(z)(x-\bar{x})(p-\bar{p}).
\end{eqnarray}
\end{subequations}
An explicit expression for $f_{G}(z)$ in terms of these moments is 
\begin{equation}
f_{G}(z) = \frac{1}{2\pi\sqrt{\det\sigma}}\exp\left[\frac{1}{2}(z - \bar{z})^{T}\cdot\sigma^{-1}\cdot(z-\bar{z})\right],
\end{equation}
where $z$ is a vector in phase space and $\sigma$ is the covariance matrix:
\begin{equation}
z=\left(\begin{array}{cc} x \\ p \end{array}\right),\qquad
\sigma = \left(\begin{array}{cc}\sigma_{x}^{2} & \sigma_{xp} \\\sigma_{xp} & \sigma_{p}^{2}\end{array}\right).
\end{equation}
Lastly, the relative entropy between two Gaussian distributions, $f_{G}(z)$ and $g_{G}(z^{*})$, is (see Eq.~\ref{eq:kpv3})
\begin{equation}\label{eq:rel_en}
\begin{split}
D\left(f_{G}(z)||g_{G}(z^*)\right) =& -1 + \frac{1}{2}\left[\ln\left(\frac{\det\sigma_{g}}{\det\sigma_{f}}\right)+\mathrm{Tr}\left(\sigma_{g}^{-1}\cdot\sigma_{f}^{*}\right)\right] \\ 
&+ \frac{1}{2}(\bar{z}_{f}^{*}-\bar{z}_{g})^{T}\cdot\sigma_{g}^{-1}\cdot(\bar{z}_{f}^{*}-\bar{z}_{g}),
\end{split}
\end{equation}
where $\sigma^{*}_{xp}=-\sigma_{xp}$ and all other elements of $\sigma^*$ are unaltered \cite{Ihara}.

\subsection{Hamiltonian Dynamics}\label{sec:det}

We now consider the case in which the system is thermally isolated from the reservoir after the initial equilibration stage, and thereafter evolves under Hamilton's equations as the harmonic well is translated leftward or rightward,
\begin{subequations}\label{eq:hd_eqs}
\begin{eqnarray}
\dot{x}_{F}  &=& p_F/m, \qquad\dot{p}_{F} =  -k(x_{F}-ut), \\
\dot{x}_{R}  &=&  p_R/m, \qquad\dot{p}_{R} =  -k(x_{R}-u\tau + ut).
\end{eqnarray}
\end{subequations}

To solve for the evolution of the phase space densities $\rho_F$ and $\rho_R$,
we observe that since the initial density is Gaussian (Eqs.~\ref{eq:init_den}), and the equations of motion are linear (Eqs.~\ref{eq:hd_eqs}),
the distribution remains Gaussian for all times \cite{Kubo}.
Thus, we need only determine the means and (co)variances as functions of time.
From Eqs.~\ref{eq:init_den} we have the initial means
\begin{subequations}
\label{eq:hd_ic_all}
\begin{equation}\label{eq:hd_ic}
\bar{x}_{F}(0) = 0,\quad\bar{p}_{F}(0) = 0,\quad\bar{x}_{R}(0) = u\tau,\quad\bar{p}_{R}(0) = 0,
\end{equation}
and the initial variances
\begin{equation}\label{eq:hd_var_ic}
\sigma_x^2(0)=\frac{m}{\beta},\quad\sigma_p^2(0)=\frac{1}{\beta k},\quad\sigma_{xp}^2(0)=0,
\end{equation}
\end{subequations}
which are the same for the forward and reverse processes.
Solving Eqs.\ \ref{eq:hd_eqs} leads to a set of linear equations for the positions and momenta in terms of their initial conditions.  Combining these solutions with Eqs.\ \ref{eq:gaussian} and \ref{eq:hd_ic_all} leads to the solutions
\begin{subequations}\label{eq:hd_moments}
\begin{align}
\label{eq:hd_means}
\bar{x}_{F}(t) &= u\bigg[t-\frac{\sin(\omega t)}{\omega}\bigg], &\bar{p}_{F}(t) &= mu\big[1-\cos(\omega t)\big],\\
\label{eq:hd_means2}
\bar{x}_{R}(t) &= u\bigg[\tau-t+\frac{\sin(\omega t)}{\omega}\bigg], &\bar{p}_{R}(t) &= mu\big[\cos(\omega t)-1\big],
\end{align}
\begin{equation}
\label{eq:hd_var}
\sigma_x^2(t) = \frac{m}{\beta},\quad\sigma_p^2(t) = \frac{1}{\beta k},\quad\sigma_{xp}(t)=0,
\end{equation}
\end{subequations}
where $\omega^{2}=k/m$. 

We can now explicitly verify Eq.\ \ref{eq:kpv2}.
The relative entropy between the forward phase space density $\rho_F(z,t)$ and the reverse phase space density $\rho_R(z^*,\tau-t)$ is determined by plugging Eqs.\ \ref{eq:hd_moments} into Eq.\ \ref{eq:rel_en},
\begin{equation}
D(\rho_{F}(z,t)||\rho_{R}(z^{*},\tau-t))= \beta mu^{2}\big(1-\cos(\omega \tau)\big).
\end{equation}
Comparing this with
\begin{equation}
\beta\langle W_{\rm diss}\rangle = \beta mu^{2}\big(1-\cos(\omega \tau)\big),
\end{equation}
computed from Eqs.\ \ref{eq:hd_means} and \ref{eq:avg_work}, we find the predicted result $\beta\langle W_{{\rm diss}}\rangle=D(\rho_{F}||\rho_{R})$.
 
 \subsection{Overdamped Langevin Dynamics}\label{sec:od}
 
We now imagine that the system remains in contact with the thermal reservoir throughout the process, and we will use Langevin dynamics to model the presence of the reservoir.
As discussed in Section~\ref{sec:rev}, Eq.~\ref{eq:kpv2} should apply as a strict inequality in this case.
 
In this section we consider the overdamped limit, in which the momentum effectively equilibrates instantaneously, and as a result the momentum does not contribute to the relative entropy.
We therefore focus on the position dynamics.
For the forward process the Fokker-Planck equation for $\rho_F(x,t)$ is
 \begin{equation}\label{eq:od_fp}
 \frac{\partial}{\partial t}\rho_F(x,t) = \frac{k}{\gamma}\frac{\partial}{\partial x}\left[(x-ut)\rho_F(x,t)\right]+\frac{1}{\gamma\beta}\frac{\partial^{2}}{\partial x^{2}}\rho_F(x,t),
 \end{equation}
where $\gamma$ is the friction coefficient.

To solve Eq.\ \ref{eq:od_fp}, we recognize that an initially Gaussian distribution will remain Gaussian for all time under the evolution of Eq.\ \ref{eq:od_fp}, as can be checked by substitution (see for example  \cite{vanZon2003a, Mazonka1999}).  
Thus, $\rho_F(x,t)$ is Gaussian with mean and variance,
 \begin{equation}\label{eq:od_x_sol}
 \bar{x}_{F}(t) = ut-\frac{\gamma u}{k}\left(1-e^{-kt/\gamma}\right),\quad\sigma_F^2=\frac{1}{\beta k}.
 \end{equation}
 For the reverse process we replace $t$ with $\tau-t$ in Eq.~\ref{eq:od_fp}.  The solution is a Gaussian distribution with mean and variance,
  \begin{equation}
 \bar{x}_{R}(t) = u(\tau -t)+\frac{\gamma u}{k}\left(1-e^{-k(\tau-t)/\gamma}\right),\quad\sigma_R^2=\frac{1}{\beta k}.
 \end{equation}
 
 We now use these results to calculate the relative entropy and average dissipated work.
 Using Eq.~\ref{eq:rel_en} we obtain
 \begin{equation}\label{eq:od_rel}
 \begin{split}
D(\rho_{F}(x,t)||\rho_{R}(x,\tau-t)) = & \frac{2\beta\gamma^{2}u^{2}}{k}\bigg[1 \\
& -e^{-k\tau/2\gamma}\cosh\left(\frac{k}{\gamma}\left(\frac{\tau}{2}-t\right)\right)\bigg]^{2}.
\end{split}
 \end{equation}
The average dissipated work is obtained by substituting Eq.~\ref{eq:od_x_sol} into Eq.~\ref{eq:avg_work} and evaluating the integral:
\begin{equation}
\label{eq:od_wdiss}
\beta\langle W_{\rm diss}\rangle = \beta\gamma u^{2}\left[\tau - \frac{\gamma}{k}\left(1-e^{-k\tau/\gamma}\right)\right].
\end{equation}

\begin{figure}[htb]
\centering
\includegraphics[scale=0.45]{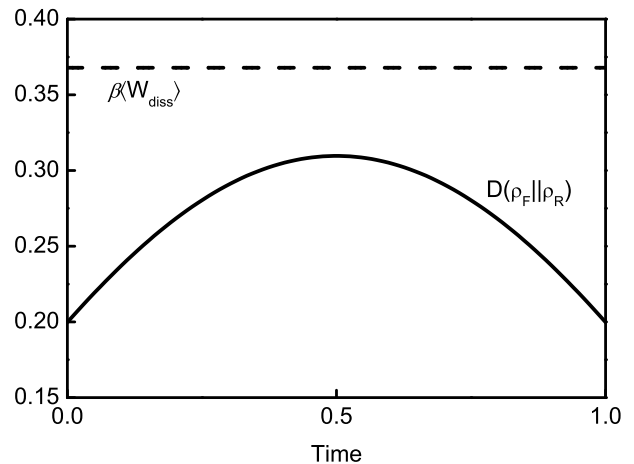}
\caption{Comparison of $\beta\langle W_{\rm diss}\rangle$ (dashed line) and $D(\rho_{F}||\rho_{R})$ (solid line) in the overdamped limit.  All parameters ($k$, $\gamma$, $u$ and $\beta$) have been set to one in their respective units.}
\label{fig:od}
\end{figure}

Verifying Eq.\ \ref{eq:kpv2} requires demonstrating that the average dissipated work is always greater than the relative entropy for any values of system parameters.
To begin, we note that $\beta\langle W_{\mathrm{diss}}\rangle$ does not depend on $t$ and $D(\rho_F||\rho_R)$ obtains its maximum value $D_{\textrm{max}}$ at time $t=\tau/2$.
Combining Eqs.\ \ref{eq:od_rel} and \ref{eq:od_wdiss}, we get
\begin{equation}\label{eq:od_diff}
\begin{split}
\beta\langle W_{\textrm{diss}}\rangle-D(\rho_F||\rho_R)&\ge\beta\langle W_{\textrm{diss}}\rangle-D_{\textrm{max}} \\
&=\frac{\beta\gamma^2 u^2}{k}\left[\zeta-\left(3-e^{-\zeta/2}\right)\left(1-e^{-\zeta/2}\right)\right] \\
&\ge0,
\end{split}
\end{equation}
where $\zeta=k\tau/\gamma$ is the scaled time. 
Using introductory calculus techniques, it is easy to verify that the bracketed quantity in Eq.\ \ref{eq:od_diff} is non-negative for any $\zeta\ge0$.
As an example, we plot $\beta\langle W_{\textrm{diss}}\rangle$ and $D(\rho_F||\rho_R)$ in Fig.\ \ref{fig:od} as functions of time for $k$, $\gamma$, $u$ and $\beta$ all set to one.

In the quasi-static limit, namely $u\to 0$ with $u\tau$ fixed, both the dissipation and relative entropy approach zero, although at different rates.  
Specifically, from Eqs.~\ref{eq:od_rel} and \ref{eq:od_wdiss}, $\beta\langle W_{\textrm{diss}}\rangle\sim u$ and $D(\rho_F||\rho_R)\sim u^2$.
This limiting behavior can be justified on general grounds since we are considering continuous Markovian stochastic processes \footnote{S. Vaikuntanathan, J. Horowitz, S. Rahav, and C. Jarzynski, {\it unpublished}.}.
A heuristic argument is as follows: consider such a continuous Markovian stochastic process perturbed by varying an external parameter $\lambda$ according to a specified protocol $\lambda(t)$.
In the quasi-static limit, i.e.\ $\dot\lambda\sim\epsilon\ll1$, the phase space density is approximately 
\begin{equation}\label{eq:limit}
P(z,t)\sim P_{\lambda(t)}^{ss}(z)+\epsilon\delta P(z,t),
\end{equation}
where $P_\lambda^{ss}(z)$ is the unique stationary state with fixed external parameter $\lambda$ and $\delta P(z,t)$ is the first order correction to the phase space density \footnote{Equation \ref{eq:limit} can be derived from the Fokker-Planck equation by employing a two-time scale analysis.}.  
Combining Eq.\ \ref{eq:limit} with the definitions of average dissipated work (cf. Eq.\ \ref{eq:work})  and relative entropy, and taking the limit $\epsilon\to 0$, leads to $\beta\langle W_{\textrm{diss}}\rangle\sim\epsilon$ and $D(\rho_F||\rho_R)\sim\epsilon^2$.
In the above model, $\lambda(t)=ut$ with $\epsilon=u$.

\subsection{Full Phase Space Langevin Dynamics}\label{sec:full_lang}
 
In this section we analyze the same stochastic system without assuming the overdamped limit.
(We do assume that the motion is not critically damped, i.e. $\gamma^{2} \ne 4mk$.)  Using the same notation as the previous section (\ref{sec:od}) the Fokker-Planck equation for the phase space density of the forward process is
 \begin{equation}\label{eq:ps_fokker}
 \begin{split}
 \frac{\partial}{\partial t}\rho_{F}(z,t) = &-\frac{p}{m}\frac{\partial}{\partial x}\rho_{F}(z,t) \\
 &+ \frac{\partial}{\partial p}\left[\left(k(x-ut)+\frac{\gamma p}{m}\right)\rho_{F}(z,t)\right] \\
 &+\frac{\gamma}{\beta}\frac{\partial^{2}}{\partial p^{2}}\rho_{F}(z,t).
 \end{split}
 \end{equation} 

\begin{figure}[tb]
\centering
\includegraphics[scale=0.45]{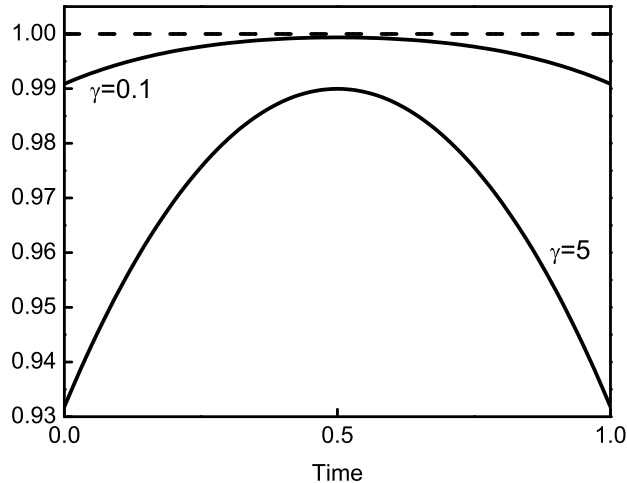}
\caption{Comparison of $\beta\langle W_{\rm diss}\rangle/D(\rho_{F}||\rho_{R})$ for various values of the friction coefficient $\gamma$;
the solid lines are, from bottom to top, $\gamma = 5$ and $\gamma=0.1$.  The dashed line represents Hamiltonian dynamics ($\gamma=0$) with $\beta\langle W_{\rm diss}\rangle=D(\rho_{F}||\rho_{R})$.   All other system parameters ($m$, $k$, $u$ and $\beta$) have been set to one in their respective units.
}
\label{fig:ps}
\end{figure}

While Eqs.\ \ref{eq:ps_fokker} can be solved exactly, the solution is complicated and unilluminating. 
An analytic solution is presented in Appendix \ref{appendix}; here, we illustrate Eq.\ \ref{eq:kpv2} by plotting the ratio $\beta\langle W_{\rm diss}\rangle/D(\rho_{F}||\rho_{R})$  in Fig.\ \ref{fig:ps} for various values of the friction coefficient $\gamma$.  
Roughly speaking, $\gamma$ measures the coupling between the system and the reservoir: for larger $\gamma$, the reservoir and system interact more strongly; whereas, when $\gamma\to 0$, the reservoir and the system decouple, with the result that the system evolves independently of the reservoir, under deterministic Hamiltonian dynamics.
Figure \ref{fig:ps} clearly shows that as $\gamma$ decreases, the relative entropy approaches the average dissipated work.
Decreasing $\gamma$ weakens the interaction between the system and  the reservoir, resulting in less microscopic information leaking into the reservoir variables.
The result is more information is contained in the system variables, reflected by an increase in the relative entropy.
The flow of microscopic information between system and reservoir degrees of freedom and its relation to dissipation has previously been discussed in the context of other models \cite{Gomez-Marin2007,Gomez-Marin2008a}.

 \section{Two-Time Phase Space Densities}\label{sec:two_time}
 
In the previous sections, the phase space densities were evaluated at one particular time.
For Hamiltonian dynamics this is enough information to determine the average dissipated work.  In contrast, when the dynamics are stochastic the phase space densities alone do not contain enough information to determine the average dissipated work. In this section we address this issue by investigating the implications of specifying the phase space density at two times.  Let  $P(z_{1},t_{1}; z_{0,}t_{0})$ be the ``two-time'' probability for the system to be at $z_{0}$ at time $t_{0}$ {\it and} $z_{1}$ at a later time $t_{1}$.  We will explicitly demonstrate that two-time distributions provide a better bound for the average dissipated work: $\beta\left\langle W_{\textrm{diss}}\right\rangle\ge D(P_{F}||P_{R})\ge D(\rho_{F}||\rho_{R})$.
 
For clarity we consider the overdamped limit as in Section \ref{sec:od}.  The goal is to determine the pairwise probability $P(x_{1},t_{1}; x_{0},t_{0})$.  
For the forward process, we can decompose the joint probability distribution into a conditional probability to be at $x_{1}$ at $t_{1}$ {\it given} the system was at $x_{0}$ at $t_{0}$, $\rho_{F}(x_{1},t_{1}|x_{0},t_{0})$, and a one-time probability $\rho_{F}(x_{0},t_{0})$, i.e.\ $P_{F}(x_{1},t_{1}; x_{0},t_{0})=\rho_{F}(x_{1},t_{1}|x_{0},t_{0})\rho_{F}(x_{0},t_{0})$.
The expression for $\rho_{F}(x_{0},t_{0})$ was determined in Section~\ref{sec:od} (see Eq.~\ref{eq:od_x_sol}).  The conditional probability is computed by solving the Fokker-Planck equation (see Eq.~\ref{eq:od_fp}) with the initial condition $\delta(x-x_{0})$ at $t_{0}$, i.e.~the initial density is Gaussian with $\bar x(t_{0})=x_{0}$ and $\sigma_{x}^{2}(t_{0})=0$. At time $t_{1}$, the mean and variance for the forward process conditional probability are
\begin{subequations}
\begin{equation}
\begin{split}
\bar x(t_{1}) = &e^{-k(t_{1}-t_{0})/\gamma}x_{0}+u\left(t_{1}-t_{0}e^{-k(t_{1}-t_{0})/\gamma}\right) \\
&-\frac{\gamma u}{k}\left(1-e^{-k(t_{1}-t_{0})/\gamma}\right),
\end{split}
\end{equation}
\begin{equation}
\sigma_{x}^{2}(t_{1}) = \frac{1}{\beta k}\left(1-e^{-2k(t_{1}-t_{0})/\gamma}\right).
\end{equation}
\end{subequations}
 The pairwise distribution is the product of Gaussian distributions; as such, it is Gaussian as well.  A simple, yet lengthy calculation shows that the means and variances of the pairwise distribution for the forward process are
 \begin{subequations}
 \begin{equation}
 \bar{\mathbf{x}}_{F}=\left(\begin{array}{c}\bar{x}_{0} \\ \bar{x}_{1}\end{array}\right)
 = \left(\begin{array}{c}
 ut_{0}-\frac{\gamma u}{k}\left(1-e^{-kt_{0}/\gamma}\right) \\
 ut_{1}-\frac{\gamma u}{k}\left(1-e^{-kt_{1}/\gamma}\right)
 \end{array}\right)
\end{equation}
\begin{equation}
\sigma_{F} = \left(\begin{array}{cc}\sigma_{x_{0}}^{2} & \sigma_{x_{0}x_{1}} \\\sigma_{x_{0}x_{1}} & \sigma_{x_{1}}^{2}\end{array}\right)
=\frac{1}{\beta k}\left(\begin{array}{cc} 1 & e^{-k(t_{1}-t_{0})/\gamma} \\
e^{-k(t_{1}-t_{0})/\gamma} & 1
\end{array}\right)
\end{equation}
\end{subequations}
 Repeating the above calculation for the reverse joint distribution leads to similar results.  
 
 The next step is to determine the relative entropy.  We are interested in comparing the pairwise probability for the forward process $P_{F}(x_{1},t_{1}; x_{0},t_{0})$, with the {\it time-reversed} pairwise probability for the reverse process $P_{R}(x_{0},\tau-t_{0};x_{1},\tau-t_{1})$.  
 If we allow $t_{0}$ and $t_{1}$ to be arbitrary, then from Eq.~\ref{eq:rel_en} the relative entropy is
 \begin{equation}\label{eq:two_time_relEnt}
 \begin{split}
 D(P_{F}||P_{R})=&\frac{2\beta\gamma^{2}u^{2}/k}{1-e^{-2k|t_{1}-t_{0}|/\gamma}} \\
 &\times\left(\alpha_{0}^{2}+\alpha_{1}^{2}-2e^{-k|t_{1}-t_{0}|/\gamma}\alpha_{0}\alpha_{1}\right)
 \end{split}
 \end{equation}
 where
 \begin{equation}
 \alpha_{i}=1-e^{-k\tau/2\gamma}\cosh\left(\frac{k}{\gamma}\left(\frac{\tau}{2}-t_{i}\right)\right),\quad i=0,1.
 \end{equation}
 \begin{figure}
\includegraphics[scale=0.46]{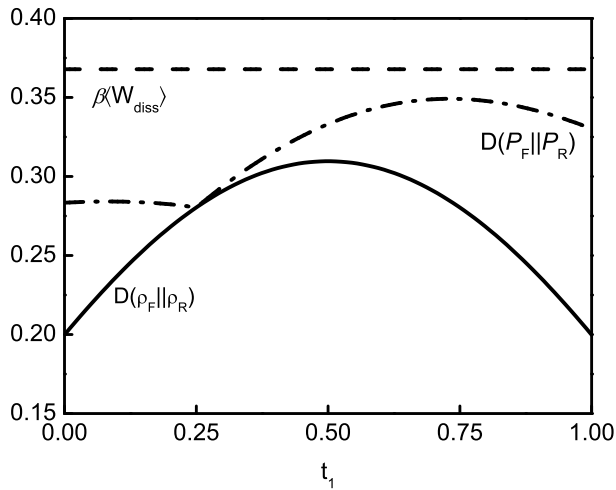}
\caption{Comparison of $D(P_{F}||P_{R})$ (dot-dashed line), $D(\rho_{F}||\rho_{R})$ (solid line) and  $\beta\langle W_{\rm diss}\rangle$ (dashed line) in the overdamped limit.   All system parameters have been set to one in their respective units.}
\label{fig:two_time}
\end{figure}

Figure \ref{fig:two_time} compares the ``two-time'' relative entropy $D(P_{F}||P_{R})$, ``one-time'' relative entropy $D(\rho_{F}||\rho_{R})$, and the average dissipated work.  For $D(P_{F}||P_{R})$, we fix $t_{0}=\tau/4$.
We see that
$\beta\langle W_{\rm diss} \rangle \ge D(P_{F}||P_{R}) \ge D(\rho_{F}||\rho_{R})$ for all times $t_1$.
Thus, by specifying an additional time in our probability distributions we have improved the bound in Eq.\ \ref{eq:kpv2}. 

This result is an instance of the coarse-graining procedure outlined in Section \ref{sec:rev} (see the discussion below Eq.\ \ref{eq:pathSpace}), where the path space probability distribution $\mathcal{P}_F(\gamma^F)$ has been projected down onto a two-time phase space density.
Projecting onto a two-time phase space density eliminates less microscopic information than projecting onto a one-time phase space density;
hence, the two-time relative entropy is greater than the one-time relative entropy.

In a related independent analysis, Gomez-Marin {\it et al.}\ \cite{Gomez-Marin2007, Gomez-Marin2008b} used the present model
to investigate the behavior of the ``$N$-time'' relative entropy $D_N$.
By evaluating $D_N$ at $N$ equally spaced times, i.e.\ $t_{i+1}-t_{i}=\tau/N$, they were able to demonstrate that $\beta\langle W_{\mathrm{diss}}\rangle-D_N\sim1/N^2$ as $N\to\infty$.
We note that our two-time relative entropy agrees with theirs: choosing $t_1=\tau$ and $t_0=0$ in Eq.\ \ref{eq:two_time_relEnt},  we recover Eq.\ 30 of Ref. \cite{Gomez-Marin2007} and Eq.\ 22 (with $n=1$) of Ref.\ \cite{Gomez-Marin2008b}.

 \section{Conclusion}

The average energy dissipated by a thermodynamic process is related to how distinguishable the process is from its time-reversal, i.e.\ how well one can discern the ``arrow of time".
This paper provides a pedagogical illustration of this idea using an exactly solvable model.
For deterministic Hamiltonian systems, complete knowledge of the forward and reverse phase space density at any {\it one} time is enough to determine the dissipation.
This is not true for stochastic systems.
Calculating the dissipation from microscopic information in stochastic systems requires knowledge of the entire evolution of the system, both for the forward and reverse processes.
Partial knowledge gives only a lower bound on the dissipation.
We have seen that the tightness of the lower bound is correlated with the amount of information known about the system's evolution.
Specifically, when the system is loosely coupled to a reservoir, very little information is lost to the reservoir.
As a result the relative entropy between the system's forward and reverse phase space densities reasonably approximates the dissipation in this limit.
The lower bound can also be tightened using a multi-time relative entropy where the microscopic state of the system is specified many times along the system's trajectory.

\acknowledgments
We thank S. Vaikuntanathan and A. Ballard for a critical reading of this manuscript and the University of Maryland for their generous support.

\appendix
\section{Phase Space Density for Section \ref{sec:full_lang}}\label{appendix}

In this appendix, we solve Eq.\ \ref{eq:ps_fokker} for the full phase space density of Section \ref{sec:full_lang}.
To begin, the coefficients of Eq.\ \ref{eq:ps_fokker} are linear in $x$ and $p$.  
Consequently, as in the previous sections, an initial Gaussian distribution will remain Gaussian.
Again we must determine the means, $\bar{x}$ and $\bar{p}$, and covariance matrix $\sigma$ as functions of time. 
Combining Eq.~\ref{eq:ps_fokker} with the derivatives of Eqs.~\ref{eq:gaussian} we find the equations of motion for the means and variances of the forward process
\begin{subequations}\label{eq:ps}
 \begin{equation}\label{eq:ps_means}
 \frac{d\bar{x}_{F}}{dt} = \frac{\bar{p}_{F}}{m}, \quad
 \frac{d\bar{p}_{F}}{dt} =  -k(\bar{x}_{F}-ut)-\frac{\gamma}{m}\bar{p}_{F},
 \end{equation}
 \begin{equation}\label{eq:ps_var}
 \begin{split}
&\frac{d\sigma^2_{x}}{dt} = \frac{2}{m}\sigma_{xp},\quad
 \frac{d\sigma^2_{p}}{dt} =  -\frac{2\gamma}{m}\sigma^2_p - 2k\sigma_{xp}+\frac{2\gamma}{\beta},\\ 
 &\qquad\qquad\frac{d\sigma_{xp}}{dt} =  -\frac{\gamma}{m}\sigma_{xp} - k\sigma_x^2+\frac{1}{m}\sigma^2_p. 
\end{split}
\end{equation}
\end{subequations}
The reverse process phase space density is described by a similar set of equations.

The solutions of Eqs.\ \ref{eq:ps_means} and their reverse process counterparts are
\begin{subequations}
\begin{eqnarray}
\bar{x}_{F}(t) &=& Ae^{r_{+}t}+Be^{r_{-}t}+u(t-\gamma/k), \\
\bar{p}_{F}(t) &=& mr_{+}Ae^{r_{+}t}+mr_{-}Be^{r_{-}t}+mu, \\
\bar{x}_{R}(t) &=& -Ae^{r_{+}t}-Be^{r_{-}t}+u(\tau-t+\gamma/k), \\
\bar{p}_{R}(t) &=& -mr_{+}Ae^{r_{+}t}-mr_{-}Be^{r_{-}t}-mu,
\end{eqnarray}
\end{subequations}
where
\begin{equation}
r_{\pm}=\frac{-\gamma\pm\sqrt{\gamma^{2}-4mk}}{2m}
\end{equation}
and
\begin{equation}
\left(\begin{array}{c}A \\B\end{array}\right)=\frac{-mu}{\sqrt{\gamma^{2}-4mk}}\left(\begin{array}{c}\gamma r_{+}/k+1 \\ -\gamma r_{-}/k-1\end{array}\right)
\end{equation}
are determined by the initial conditions (Eq.\ \ref{eq:hd_ic}). 
The solutions to Eq.\ \ref{eq:ps_var} with initial conditions given by Eq.\ \ref{eq:hd_var_ic} are
\begin{equation}
\sigma_x^2(t) = \frac{m}{\beta},\qquad\sigma_p^2(t)=\frac{1}{\beta k},\qquad\sigma_{xp}(t)=0.
\end{equation}
Using Eq.~\ref{eq:avg_work}, the average dissipated work is 
\begin{equation}
\left\langle W_{\mathrm{diss}}\right\rangle=\gamma u^{2}\tau+uk\left[\frac{A}{r_{+}}\left(1-e^{r_{+}t}\right)+\frac{B}{r_{-}}\left(1-e^{r_{-}t}\right)\right].
\end{equation}
From Eq.~\ref{eq:rel_en}, the relative entropy is
\begin{widetext}
\begin{equation}
\begin{split}
D(\rho_{F}||\rho_{R}) = \frac{1}{2}\bigg\{&\beta k\left[A\left(e^{r_+ t}+e^{r_+ (\tau-t)}\right)+B\left(e^{r_- t}+e^{r_- (\tau-t)}\right)\right]^2 \\
&+ \beta m\left[r_+A\left(e^{r_+ t}-e^{r_+ (\tau-t)}\right)+r_-B\left(e^{r_- t}-e^{r_- (\tau-t)}\right)\right]^2\bigg\}.
\end{split}
\end{equation}
\end{widetext}

\bibliographystyle{apsrev} 
\bibliography{FluctuationTheory,PhysicsTexts}

\end{document}